# Motion of Coloured Particles in Solitons of the $O(3)$ Non-Linear Model

**Rafael M. Fernandes**[*]
*Instituto de Física "Gleb Wataghin", Universidade Estadual de Campinas, 13083-970, Campinas, SP, Brazil*
*E-mail:* `rafaelmf@ifi.unicamp.br`

**Patricio S. Letelier**
*Departamento de Matemática Aplicada, Instituto de Matemática, Estatística e Computação Científica, Universidade Estadual de Campinas, 13081-970, Campinas, SP, Brasil*
*E-mail:* `letelier@ime.unicamp.br`

We propose a consistent expression for the relativistic quadri-force referring to the classical interaction between a coloured particle and a scalar field multiplete. General aspects of the resultant equations of motion are discussed as well as a specialisation to the case of a $SU(2)$ particle in the presence of a soliton configuration of the non-linear $O(3)$ model with unit topological charge. Analytical studies regarding the stability and asymptotic behaviour of this system are made and extended to the case of arbitrary topological charge. Since the non-linear $O(3)$ model describes an isotropic ferromagnet, we also investigate a possible physical interpretation of such a system.

---

[*]Speaker.

## 1. Introduction

The interaction between a particle and a scalar field when both have a non-Abelian internal space is still a matter of investigation. In one of the first works regarding the subject, Wong [1] proposes an expression, derived from the Dirac equation, for the classical interaction between a particle with $SU(2)$ internal symmetry group and a vector field, namely, the Yang-Mills gauge field. However, there was no reference to the interaction with a scalar field. Fehér [2] generalises Wong's procedure to five dimensions in an attempt to take into account the existence of a scalar field, but as already pointed out by Azizi [3], the generalisation seems too much unnatural. In his own work, Azizi makes use of a fifth dynamical dimension in order to generalise again Wong's procedure, but the expression he achieves for the quadri-force, although consistent in a Newtonian limit, is inconsistent in a relativistic frame, since it is not orthogonal to the quadri-velocity of the particle.

Therefore, we propose a consistent way to obtain the expression of the quadri-force describing the interaction of a particle and a scalar field by looking directly for its possible forms compatible with Special Relativity dynamics. The temporal evolution of the internal vector, which is the other relevant equation of motion, is achieved considering a natural extension of Wong's expression. We investigate the expressions proposed in a general scenario and then specialise to the case in which the scalar field is a soliton type configuration of the non-linear $O(3)$ model. For this particular system, we make analytical studies of its stability and asymptotic behaviour, complementing with numerical evaluation of some orbits and Lyapunov exponents.

In Section 2 we explicitly state the equations of motion due to the particle's interaction with a scalar field multiplete in the observer frame and investigate their Newtonian limit and the question if they can be expressed as Hamilton equations. In Section 3, we take the expressions of the soliton field configurations of the non-linear $O(3)$ model and apply to the previous equations of motion. Hence, we study the stability and the asymptotic behaviour of the resultant system in Section 4 and also calculate numerically some Lyapunov exponents. Finally, in Section 5, based on the fact that the considered model describes an isotropic ferromagnet, we make a physical interpretation of the system.

## 2. Quadri-force and equations of motion

To construct the quadri-force, first of all, we have to respect the basic relativistic property that quadri-acceleration must be orthogonal to quadri-velocity, since $u^\mu u_\mu = 1$. If, in addition, we impose a minimum coupling between the internal vector of the particle (its generalised charge) and the scalar field, we can settle that there are basically two possible forms for the quadri-force: one is achieved by using the totally antisymmetric tensor $\varepsilon^{\mu\nu\lambda\rho}$, namely:

$$\overline{K}^\mu = q\varepsilon^{\mu\nu\lambda\rho}\frac{dx_\nu}{d\tau}\phi^a_{,\lambda}I^a W_\rho, \tag{2.1}$$

in which $q$ is the coupling constant, $\phi^a$ the scalar field, $I^a$ the internal vector of the particle, $x^\mu(\tau)$ its world line and $\tau$ its proper time. Let us analyse the undetermined term $W_\rho$: it cannot be related to the scalar field, since we are assuming minimum coupling, and cannot be quadri-velocity either. Hence, the remaining possibilities for it are the quadri-acceleration or an external field, which are rather artificial for the motion of a single particle and will not be considered in the scope of this paper.

The other possible form for the quadri-force is written in terms of the usual tensor that projects any vector in the subspace orthogonal to the quadri-velocity:

$$K^\mu = m\frac{d^2x^\mu}{d\tau^2} = q\left(\eta^{\mu\nu} - \frac{dx^\mu}{d\tau}\frac{dx^\nu}{d\tau}\right)\phi^a_{,\nu}I^a, \qquad (2.2)$$

in which $\eta^{\mu\nu}$ is the Minkowskian metric. This expression is quadratic in the velocity and does not show any problem at a first sight. Moreover, it is very similar to the expression Barut [4] considers for a scalar field with no internal symmetry.

The other equation of motion that must be constructed is the one involving the precession of the internal vector. Considering the internal particle's space to be the $SU(2)$ symmetry group (in which case $I^a$ is called the isospin of the particle) and then making the simplest generalisation of Wong's proposal by replacing the vector field $A^a_\mu$ by $\phi^a_{,\mu}$, we are lead to:

$$\frac{dI^a}{d\tau} + q\varepsilon^{abc}\phi^b_{,\mu}I^c\frac{dx^\mu}{d\tau} = 0. \qquad (2.3)$$

Next, we move to the observer frame, which is related to the particle's frame by the well-known identity $dt = \gamma d\tau$. Assuming that the field does not depend upon time explicitly, we reach the following dynamical system:

$$\dot{x}^j = v^j \ ; \ \dot{v}^j = -\frac{q}{m}\gamma^{-2}\phi^a_{,j}I^a \ ; \ \dot{\vec{I}} = \left(\vec{I}\times\vec{\phi}_{,j}\right)v^j, \qquad (2.4)$$

in which we have used the convention that a observer time derivative is represented by ( ˙ ) and a vectorial notation for the internal variables. The dynamical system (2.4) has the desirable property of respecting symmetries of the field, i.e., if the latter does not depend upon a certain coordinate, then there will be no acceleration in the respective direction. We can also note that the module of the isospin is a constant of motion, whilst the energy seems not to be, since we have the equation:

$$\frac{d}{dt}\left(m\ln\frac{T}{m} + q\vec{\phi}\cdot\vec{I}\right) = q\vec{\phi}\cdot\frac{d\vec{I}}{dt}, \qquad (2.5)$$

in which $T = m\gamma$ is the total relativistic kinetic energy of the particle. Hence, it is clear that the origin of the dissipation in the system is the coupling to the isospin, which varies in time in general. However, if there is a special situation in which it is constant in time, then energy will be conserved; we observe that such situation can be achieved only if the isospin is parallel, in the internal space, to the time derivative of the field. To see that, we write equation (2.3) in the observer frame, $\dot{\vec{I}} = \vec{I}\times\dot{\vec{\phi}}$, from where we conclude that, if the internal vector is to remain constant, the time derivative of the field and, consequently, the field itself after a certain time, must point always in the same direction in the internal space. Hence, if one considers soliton type configuration of the fields, this condition will just be satisfied by zero topological charge solitons.

A last important remark to be made about (2.2) and (2.3) is when one considers scalar fields coming from gauge invariant Lagrangians (such as the Higgs fields in the Yang-Mills Lagrangian). The expressions may easily be made also gauge invariant by replacing the ordinary derivatives by the respective covariant ones.

## 3. Specialisation to the non-linear $O(3)$ model

All the formalism we have developed in the section before may now be applied to a concrete case to see the basic features of the system. We choose a scalar field multiplete with no gauge symmetry by taking a soliton type field configuration of the non-linear $O(3)$ model. The Lagrangian describing this bi-dimensional model is given by [5]:

$$L = \frac{1}{2}\partial_\mu \phi^a(x,y) \partial^\mu \phi^a(x,y), \tag{3.1}$$

subjected to the constraint $\phi^a \phi^a = 1$. The resultant equations of motion admit static soliton solutions, i.e., static field configuration with constant and localised energy $E = 4\pi Q$, where $Q$ is its topological charge. In polar coordinates, these solitons are given by:

$$\phi^1 = \frac{4r^Q \cos Q\alpha}{r^{2Q}+4} \;;\; \phi^2 = \frac{4r^Q \sin Q\alpha}{r^{2Q}+4} \;;\; \phi^3 = \frac{r^{2Q}-4}{r^{2Q}+4}. \tag{3.2}$$

Let us then analyse the motion of a coloured particle in the presence of the static field above with unit topological charge ($Q = 1$). Applying this field configuration to the equations (2.4), we have the dynamical system:

$$\begin{aligned}
\dot{x} &= v_x \;;\; \dot{y} = v_y \\
\dot{v}_x &= \frac{-4q}{m} \frac{(1-v_x^2-v_y^2)}{(x^2+y^2+4)^2}[I_1(-x^2+y^2+4) - 2I_1 xy + 4x\varepsilon\sqrt{1-I_1^2-I_2^2}] \\
\dot{v}_y &= \frac{-4q}{m} \frac{(1-v_x^2-v_y^2)}{(x^2+y^2+4)^2}[-2I_1 xy + I_2(x^2-y^2+4) + 4y\varepsilon\sqrt{1-I_1^2-I_2^2}] \\
\dot{I}_1 &= \frac{-4q}{(x^2+y^2+4)^2}\{v_x[-4xI_2 - 2xy\varepsilon\sqrt{1-I_1^2-I_2^2}] + \\
& \quad v_y[-4yI_2 + (x^2-y^2+4)\varepsilon\sqrt{1-I_1^2-I_2^2}]\} \\
\dot{I}_2 &= \frac{-4q}{(x^2+y^2+4)^2}\{v_x[4xI_1 - (-x^2+y^2+4)\varepsilon\sqrt{1-I_1^2-I_2^2}] + \\
& \quad v_y[4yI_1 + 2xy\varepsilon\sqrt{1-I_1^2-I_2^2}]\},
\end{aligned} \tag{3.3}$$

in which $\varepsilon = \pm 1$, what comes from the choice of Cartesian coordinates $(I_1, I_2, I_3)$ to describe the internal space, $I_3 = \pm\sqrt{1-I_1^2-I_2^2}$. Therefore, this choice implies that the internal space, which is actually a unitary spherical surface, is divided in two separate subspaces, namely, its hemispheres $I_3 > 0$ and $I_3 < 0$. From this fact, it would be expected that spherical coordinates would describe this space better than the Cartesian. However, the equations that result from the substitution of spherical coordinates present divergences in the poles of the internal sphere and serious numerical problems.

## 4. Stability and asymptotic behaviour

Although no analytical solutions of the equations of motion seem possible, important analytical features of the system can be carried on, such as stability and asymptotic behaviour. In what follows, we are considering $m = q = 1$. Regarding stability, we find from (3.3) that the equilibrium points lie on a 2 dimensional surface in the 4 dimensional space $x, y, I_1, I_2$, which is the 6 dimensional phase space subjected to the cuts $v_x = v_y = 0$. This surface may be parametrised in terms of $x$ and $y$ to give:

$$I_{(1,2)} = \pm \frac{4\varepsilon x_{(1,2)}}{x^2 + y^2 + 4}, \begin{cases} - \text{ if } (x,y) \in D_2 \cup \delta D_2 \\ + \text{ if } (x,y) \notin D_2 \end{cases}, \quad (4.1)$$

with $D_2 \equiv \{(x,y) \in \Re^2 | x^2 + y^2 < 4\}$. We note from (4.1) that the equilibrium condition is equivalent to $I^a = \pm \phi^a$, which could have been seen directly from the expressions (2.2) and (2.3) written as first order equations and in vector notation for the internal quantities:

$$\frac{d\vec{I}}{d\tau} + \vec{I} \times \vec{\phi}_{,\mu} u^\mu = 0 \; ; \; m\frac{du^\mu}{d\tau} = q(\eta^{\mu\nu} - u^\mu u^\nu) \vec{I} \cdot \vec{\phi}_{,\nu} \; ; \; \frac{dx^\mu}{d\tau} = u^\mu. \quad (4.2)$$

The equilibrium condition is given by $dI^a/d\tau = dx^j/d\tau = du^\mu/d\tau = 0$; the first two of them are clearly achieved for $u^j = 0$ and $u^0 = \gamma = $ cte, and the third is reduced to $\vec{I} \cdot \vec{\phi}^{,j} = 0$. Assuming now that $I^a = \pm \phi^a$, i.e., $\vec{I} \parallel \vec{\phi}$, this equation is satisfied because of the constancy of $|\vec{\phi}|$.

Once that the equilibrium points have been determined, we classify them by calculating the eigenvalues of the stability matrix. For the case in which the parametrisation of the equilibrium surface is positive for the system with $\varepsilon = 1$, i.e., the sign in (4.1) is $+$, or it is negative for the system with $\varepsilon = -1$, this matrix is given by:

$$M_1 = \begin{pmatrix} 1 & 0 & 0 & 0 & 0 & 0 \\ 0 & 1 & 0 & 0 & 0 & 0 \\ 0 & 0 & \frac{16}{(x^2+y^2+4)^2} & 0 & \frac{4(x^2-y^2+4)}{(x^2+y^2+4)(x^2+y^2-4)} & \frac{8xy}{(x^2+y^2+4)(x^2+y^2-4)} \\ 0 & 0 & 0 & \frac{16}{(x^2+y^2+4)^2} & \frac{8xy}{(x^2+y^2+4)(x^2+y^2-4)} & \frac{4(y^2-x^2+4)}{(x^2+y^2+4)(x^2+y^2-4)} \\ \frac{-4(x^2-y^2+4)}{(x^2+y^2+4)^2} & \frac{-8xy}{(x^2+y^2+4)^2} & 0 & 0 & 0 & 0 \\ \frac{8xy}{(x^2+y^2+4)^2} & \frac{4(y^2-x^2+4)}{(x^2+y^2+4)^2} & 0 & 0 & 0 & 0 \end{pmatrix},$$

and its eigenvalues are all doubly degenerate:

$$\{0, 1, 16/(x^2 + y^2 + 4)^2\}.$$

Hence, in this case, as all the eigenvalues are positive, the equilibrium points are repulsive and the orbits near them unstable.

Now, in the case in which the parametrisation of the equilibrium surface is negative for the system with $\varepsilon = 1$ or positive for the system with $\varepsilon = -1$, the stability matrix becomes:

$$M_2 = \begin{pmatrix} 1 & 0 & 0 & 0 & 0 & 0 \\ 0 & 1 & 0 & 0 & 0 & 0 \\ 0 & 0 & \frac{-16}{(x^2+y^2+4)^2} & 0 & \frac{4(x^2-y^2+4)}{(x^2+y^2+4)(x^2+y^2-4)} & \frac{8xy}{(x^2+y^2+4)(x^2+y^2-4)} \\ 0 & 0 & 0 & \frac{-16}{(x^2+y^2+4)^2} & \frac{8xy}{(x^2+y^2+4)(x^2+y^2-4)} & \frac{4(y^2-x^2+4)}{(x^2+y^2+4)(x^2+y^2-4)} \\ \frac{4(x^2-y^2+4)}{(x^2+y^2+4)^2} & \frac{8xy}{(x^2+y^2+4)^2} & 0 & 0 & 0 & 0 \\ \frac{-8xy}{(x^2+y^2+4)^2} & \frac{-4(y^2-x^2+4)}{(x^2+y^2+4)^2} & 0 & 0 & 0 & 0 \end{pmatrix},$$

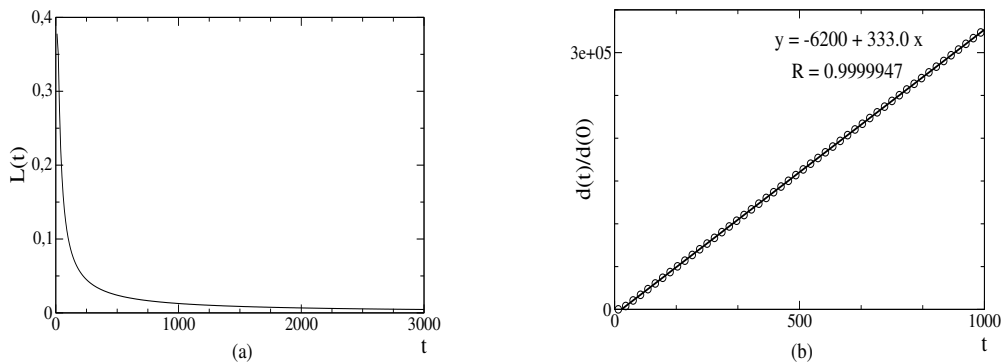

**Figure 1:** Plots referring to two initial conditions very close to the saddle point correspondent to ($x = 0.1, y = 0.2$). (a) Plot of $L(t)$, the logarithm of the quotient between the distance of the orbits in phase space after time $t$ and their initial distance divided by $t$, *versus* time. (b) Plot of the quotient between the distance of the orbits in phase space after time $t$ and their initial distance $d(t)/d(0)$ *versus* $t$, best fitted straight line to the set of points and its linear correlation coefficient $R$.

and its eigenvalues are also all doubly degenerate:

$$\{1, 0, -16/(x^2 + y^2 + 4)^2\}.$$

Therefore, as one of the eigenvalues is always positive while another one is always negative, the equilibrium points are saddle points and the orbits near them are again unstable.

Besides stability, another feature of the system that can be described analytically is its asymptotic behaviour. Observing that $\phi^a_{,i}(r \to \infty) = 0$ and then substituting it in (2.4), we get that, asymptotically, the isospin and the velocity of the particle become constant and, hence, its orbit on the $(x, y)$ plane is a straight line. Combining this with the fact that all the equilibrium points of the dynamical system are unstable, we conclude that there is no limited orbit for this system.

Therefore, it is useful to know how the distance between two initially close escaping orbits grows, to look for the possibility of chaotic behaviour. Hence, we calculated numerically the greatest Lyapunov exponent of various pairs of orbits, each with two initial conditions lying near a saddle point: one in the repulsive direction (positive eigenvector) and the other in the attractive one (negative eigenvector). All of the exponents were found to vanish.

As an example, let us consider for instance two initial conditions near the saddle point ($x = 0.1, y = 0.2$), according to the parametrisation (4.1), for the system with $\varepsilon = 1$. Figure 1 (*a*) shows that the function $L(t) = \ln(d(t)/d(0))/t$ vanishes asymptotically and, hence, the Lyapunov exponent is zero. This is confirmed by figure 1 (*b*), in which we have calculated the quotient $d(t)/d(0)$ of the Euclidean distance between the orbits, in phase space, after a period $t$ by the initial distance. The periods were equally separated in time. We have made a linear regression for the points and calculated the linear correlation of them, obtaining a value very close to unity. Hence, the quotient $d(t)/d(0)$ can be fairly approximated by a straight line, what implies in a vanishing Lyapunov exponent. We stress that the fact that the regression line does not contain the point $(0,1)$ should not cause concern as long as we are interested in the global behaviour of the quotient rather than in the local.

The analytical results before regarding stability and asymptotic behaviour can be extended to the case in which the topological charge of the field configuration given by (3.2) assumes a value $Q \geq 1$. We observe that, again, $\phi^a_{,i}(r \to \infty) = 0$, which means that the asymptotic behaviour is the same as before. Moreover, as soon as the module of the scalar field is constant for any of these configurations, the equilibrium condition is again given by $v_x = v_y = 0$ and $I^a = \pm \phi^a$ and the equilibrium points again lie on a surface. Specifically, we obtain that it is parametrised with $I_1$ and $I_2$ given by:

$$I_1 + iI_2 = \pm \frac{4\varepsilon r^Q e^{iQ}}{r^{2Q} + 4}, \begin{cases} - \text{ if } (x,y) \in D_{2^{1/Q}} \cup \delta D_{2^{1/Q}} \\ + \text{ if } (x,y) \notin D_{2^{1/Q}} \end{cases}, \tag{4.3}$$

with $D_{2^{1/Q}} \equiv \{(x,y) \in \mathfrak{R}^2 | x^2 + y^2 < 4^{1/Q}\}$.

## 5. Final remarks

The non-linear $O(3)$ model is well-known in Condensed Matter Physics as the description of an isotropic bi-dimensional ferromagnet, where the entity $\phi^a(x,y)$ is, rather than a proper scalar field, an order parameter indicating the direction of the spin in a certain position.

The topological charge of the soliton $\phi^a$ reveals possible spin configurations inside the ferromagnet; for example, while in the case $Q = 0$ the spin always point in the same direction, in the case $Q = 1$ the spin points in every direction with equal probability.

Hence, we can make a pictorial interpretation of the instability of the system. The equilibrium condition we were lead to is that the isospin of the particle has the same direction, in the internal space, of the spin in each point of the ferromagnet. As the spin configuration has unity topological charge, it points in different directions for each point in the ferromagnet. Therefore, if the particle is left in a certain point in the ferromagnet in equilibrium condition but is given a very small velocity, it will move to a point where the equilibrium condition will no longer be satisfied. Hence, to reach equilibrium, the particle will try to align its isospin with the new direction of the spin, but, as the particle is moving, this process will go on, making the isospin precessionate and the particle keep its motion, as it is exchanging energy with $\phi^a$, until it reaches a region of the space where the module of the order parameter is negligible.

The authors acknowledges FAPESP and CNPQ for financial support.